\documentclass[lettersize,journal]{IEEEtran}
\IEEEoverridecommandlockouts
\usepackage{cite}
\usepackage{amsmath,amssymb,amsfonts}
\usepackage{algorithmic}
\usepackage{graphicx}
\usepackage{textcomp}
\usepackage{xcolor}
\usepackage{gensymb}

\def\BibTeX{{\rm B\kern-.05em{\sc i\kern-.025em b}\kern-.08em
    T\kern-.1667em\lower.7ex\hbox{E}\kern-.125emX}}
\begin{document}

\title{An On-Sky Atmospheric Calibration of SPT-SLIM\\
\thanks{Identify applicable funding agency here. If none, delete this.}
}


\author{
\IEEEauthorblockN{
K.~R.~Dibert\IEEEauthorrefmark{1}\IEEEauthorrefmark{2},
M.~Adamic\IEEEauthorrefmark{3},
A.~J.~Anderson\IEEEauthorrefmark{4}\IEEEauthorrefmark{5}\IEEEauthorrefmark{1},
P.~S.~Barry\IEEEauthorrefmark{6},
B.~A.~Benson\IEEEauthorrefmark{4}\IEEEauthorrefmark{5}\IEEEauthorrefmark{1},
C.~S.~Benson\IEEEauthorrefmark{6},
E.~Brooks\IEEEauthorrefmark{1},
J.~E.~Carlstrom\IEEEauthorrefmark{5}\IEEEauthorrefmark{7}\IEEEauthorrefmark{8}\IEEEauthorrefmark{9}\IEEEauthorrefmark{1},
T.~Cecil\IEEEauthorrefmark{9},
C.~L.~Chang\IEEEauthorrefmark{9}\IEEEauthorrefmark{5}\IEEEauthorrefmark{1},
M.~Dobbs\IEEEauthorrefmark{3},
K.~Fichman\IEEEauthorrefmark{8}\IEEEauthorrefmark{5},
K.~S.~Karkare\IEEEauthorrefmark{10},
G.~K.~Keating\IEEEauthorrefmark{11},
A.~M.~Lapuente\IEEEauthorrefmark{10},
M.~Lisovenko\IEEEauthorrefmark{9},
D.~P.~Marrone\IEEEauthorrefmark{12},
J.~Montgomery\IEEEauthorrefmark{3},
T.~Natoli\IEEEauthorrefmark{5},
Z.~Pan\IEEEauthorrefmark{9}\IEEEauthorrefmark{5}\IEEEauthorrefmark{8},
A.~Rahlin\IEEEauthorrefmark{1}\IEEEauthorrefmark{5},
G.~Robson\IEEEauthorrefmark{6},
M.~Rouble\IEEEauthorrefmark{3},
G.~Smecher\IEEEauthorrefmark{13}\IEEEauthorrefmark{3},
V.~Yefremenko\IEEEauthorrefmark{9},
M.~R.~Young\IEEEauthorrefmark{4}\IEEEauthorrefmark{5},
C.~Yu\IEEEauthorrefmark{9}\IEEEauthorrefmark{5}\IEEEauthorrefmark{1},
J.~A.~Zebrowski\IEEEauthorrefmark{5}\IEEEauthorrefmark{1}\IEEEauthorrefmark{4},
C.~Zhang\IEEEauthorrefmark{14},
}
\IEEEauthorblockA{\IEEEauthorrefmark{1}Department of Astronomy and Astrophysics, University of Chicago, 5640 South Ellis Avenue, Chicago, IL, 60637, USA}
\IEEEauthorblockA{\IEEEauthorrefmark{2}Amundsen-Scott South Pole Station, South Pole, Antarctica}
\IEEEauthorblockA{\IEEEauthorrefmark{3}Department of Physics and Trottier Space Institute, McGill University, 3600 Rue University, Montreal, Quebec H3A 2T8, Canada}
\IEEEauthorblockA{\IEEEauthorrefmark{4}Fermi National Accelerator Laboratory, MS209, P.O. Box 500, Batavia, IL, 60510, USA}
\IEEEauthorblockA{\IEEEauthorrefmark{5}Kavli Institute for Cosmological Physics, University of Chicago, 5640 South Ellis Avenue, Chicago, IL, 60637, USA}
\IEEEauthorblockA{\IEEEauthorrefmark{6}School of Physics and Astronomy, Cardiff University, Cardiff CF24 3YB, United Kingdom}
\IEEEauthorblockA{\IEEEauthorrefmark{7}Enrico Fermi Institute, University of Chicago, 5640 South Ellis Avenue, Chicago, IL, 60637, USA}
\IEEEauthorblockA{\IEEEauthorrefmark{8}Department of Physics, University of Chicago, 5640 South Ellis Avenue, Chicago, IL, 60637, USA}
\IEEEauthorblockA{\IEEEauthorrefmark{9}High-Energy Physics Division, Argonne National Laboratory, 9700 South Cass Avenue., Lemont, IL, 60439, USA}
\IEEEauthorblockA{\IEEEauthorrefmark{10}Department of Physics, Boston University, 590 Commonwealth Avenue, Boston, MA, 02215, USA}
\IEEEauthorblockA{\IEEEauthorrefmark{11}Harvard-Smithsonian Center for Astrophysics, 60 Garden Street, Cambridge, MA, 02138, USA}
\IEEEauthorblockA{\IEEEauthorrefmark{12}Steward Observatory and Department of Astronomy, University of Arizona, 933 N. Cherry Ave., Tucson, AZ 85721, USA}
\IEEEauthorblockA{\IEEEauthorrefmark{13}Three-Speed Logic, Inc., Victoria, B.C., V8S 3Z5, Canada}
\IEEEauthorblockA{\IEEEauthorrefmark{14}SLAC National Accelerator Laboratory, 2575 Sand Hill Road, Menlo Park, CA, 94025, USA}}

\maketitle

\begin{abstract}
We present the methodology and results of the on-sky responsivity calibration of the South Pole Telescope Shirokoff Line Intensity Mapper (SPT-SLIM). SPT-SLIM is a pathfinder line intensity mapping experiment utilizing the on-chip spectrometer technology, and was first deployed during the 2024-2025 Austral Summer season on the South Pole Telescope. During the two-week on-sky operation of SPT-SLIM, we performed periodic measurements of the detector response as a function of the telescope elevation angle. Combining these data with atmospheric opacity measurements from an on-site atmospheric tipping radiometer, simulated South Pole atmospheric spectra, and measured detector spectral responses, we construct estimates for the responsivity of SPT-SLIM detectors to sky loading. We then use this model to calibrate observations of the moon taken by SPT-SLIM, cross-checking the result against the known brightness temperature of the Moon as a function of its phase.
\end{abstract}

\begin{IEEEkeywords}
South Pole Telescope, Line Intensity Mapping, Instrumentation
\end{IEEEkeywords}

\section{Introduction}
The South Pole Telescope Shirokoff Line Intensity Mapper (SPT-SLIM) experiment is an on-chip spectrometer camera designed  to demonstrate the use of mm-wave spectrometers for line intensity mapping \cite{Karkare_2022}.
SPT-SLIM's focal plane consists of nine pixels of dual-polarization on-chip spectrometers, with each pixel polarization coupled to a planar microstrip-based filterbank.
Each individual filter pathway terminates on a microwave kinetic inductance detector (MKID) \cite{Barry_2022, Cecil_2023}.
Of the original nine pixels, five were operational during the commissioning of SPT-SLIM and are used in this analysis.
The SPT-SLIM detectors were mounted inside a receiver that was cryogenically cooled to 150 mK during operation.
For more information on the cryogenic performance of the SPT-SLIM receiver, see \cite{Young_2025} in these proceedings.
Detectors were read out using the RF-ICE system documented in \cite{Rouble_2022}.

SPT-SLIM was deployed on the 10-meter South Pole Telescope (SPT) \cite{Carlstrom_2011} for an engineering and commissioning run in January 2025, and observed for two weeks.
During this deployment, SPT-SLIM observed with 251 active Microwave Kinetic Inductance Detectors (MKIDs) whose peak bandpass frequencies span from 110-180 GHz. 
The commissioning run included the observation of several astronomical sources, including the moon.
To extract spectra from these sources, the detectors must be calibrated by establishing a conversion from the MKID frequency shift response to an effective brightness temperature.
This is done by measuring the in-situ detector response to a known optical load.
This work uses the detector response to a set of atmospheric loads to calibrate the SPT-SLIM detectors. 
A similar atmospheric calibration method has previously been used to calibrate the superconducting spectrometer DESHIMA \cite{Takekoshi_2020}.
Though a useful check in general, atmospheric calibration is particularly necessary in this case because the fixed-temperature calibrator mounted on the secondary mirror malfunctioned during the initial commissioning of the SPT-SLIM receiver, and was therefore not operational.

During the SPT-SLIM campaign, we performed fifteen atmospheric response measurements at varying times of day and in varying weather conditions.
These consisted of a frequency sweep and noise stare at telescope elevation angles ranging from 15-80 degrees in 5 degree increments, with the telescope pointing away from the sun.
Simultaneous atmospheric opacity data from an on-site atmospheric tipping radiometer \cite{Radford_2016} was used to estimate the precipitable water vapor (PWV) during each measurement.
The PWV values were then used in the \textit{am} software \cite{am_software} to create simulated atmospheric temperature and opacity spectra for each measurement.
Using the Fourier Transform Spectrometer (FTS) data taken just prior to the placement of the SPT-SLIM receiver in the SPT cabin, we estimated the spectral bandpass for each SPT-SLIM detector, then integrated this bandpass over the atmospheric spectra generated by \textit{am} to obtain an estimated load on each detector.
We fit the detector frequency response data during each measurement to a theoretical model, and from this computed responsivity estimates for each detector.
Finally, we use the results of the detector response fits to calibrate observations of the moon performed near the end of the SPT-SLIM campaign, and to construct temperature spectra of the moon.

\section{Detector Bandpasses}

\begin{figure}[t]
\includegraphics[width=8cm, trim={0.6cm, 0.6cm, 0.6cm, 0.6cm}, clip]{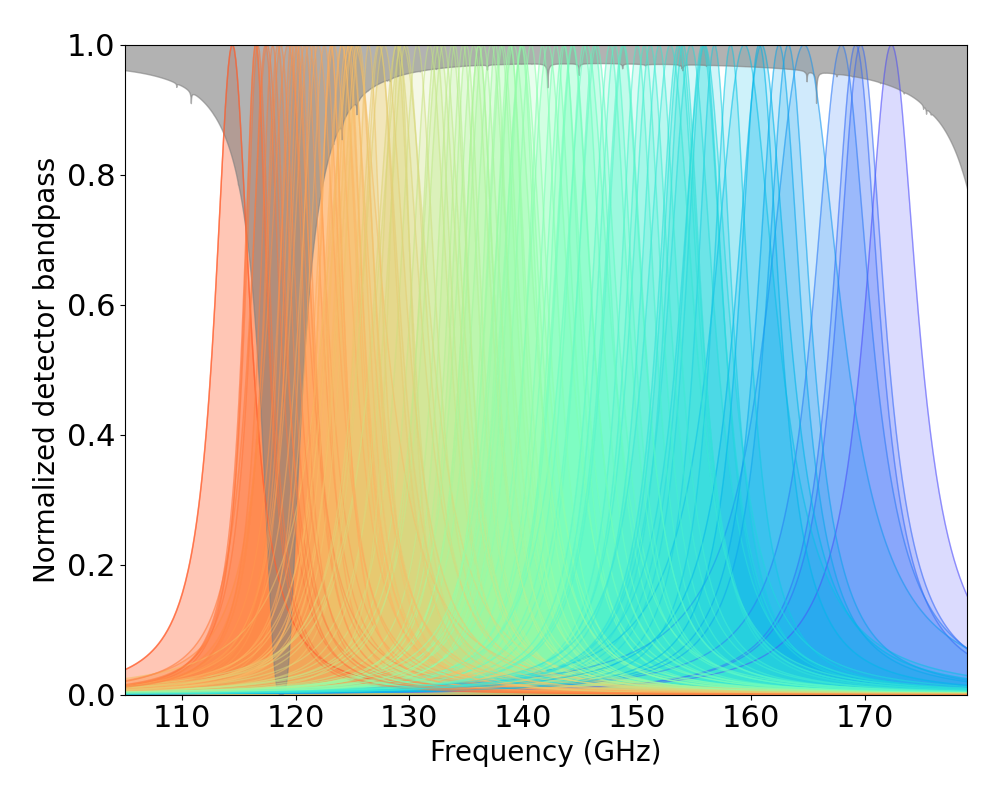}
\centering
\caption{Best fit detector bandpasses for one of the five operational SPT-SLIM filterbanks. This set of bandpasses was measured in the SPT-SLIM cryostat at the South Pole using a Fourier Transform Spectrometer (FTS) \cite{Liu_20, Pan_19}. Each position-space interferogram generated from the FTS was fit with a decaying cosine function, which was then Fourier transformed back into a frequency-space Lorentzian peak. Median atmospheric transmission for the South Pole summer is overplotted in gray.} 
\label{fig:fts}
\end{figure}

The bandpass of each detector is required to calculate its atmospheric loading.
This analysis uses detector bandpasses derived from FTS data taken at the South Pole in December 2024.
For a more in-depth analysis of SPT-SLIM bandpass measurements, see \cite{Fichman_2025} in these proceedings.
The resolution of the South Pole FTS is only 4 GHz \cite{Liu_20, Pan_19}.
Fitting a template function in the optical delay domain allows the resolution of the profile to be measured more precisely, so best-fit Lorentzian bandpasses were obtained by fitting a decaying cosine function to the position-space interferograms:

\begin{equation}
\label{eq:interf_fit}
I(z) = e^{- \sigma_0 \pi z/ R} \cos{ (2 \pi \sigma_0 z)},
\end{equation}
where $I(z)$ is the interferogram intensity, and $z$ is the optical path difference in centimeters. The fit parameters are the spectral resolution $R$ and the interferogram peak wavenumber $\sigma_0$, which corresponds to the Lorentzian peak bandpass frequency $\nu_0$ in gigahertz by $\nu_0 = \sigma_0 \times 30 \, \rm{GHz} \, \times 1 \, \rm {cm}$ (This comes from $\nu_0 = \sigma_0 \times c \times 10^{-9} \, \rm{GHz/Hz} \times  10^{2} \, \rm{cm/m}$, where $c = 3 \times 10^{8} \rm{m/s}$ is the speed of light.)
The Fourier transform of this function is the Lorentzian peak function, which is the expected form of the SPT-SLIM detector bandpass:

\begin{equation}
\label{eq:analytic_bp}
g_{\rm KID}(\nu) = \frac{\frac{1}{2} \Gamma}{ (\sigma - \sigma_0)^2 + (\frac{1}{2} \Gamma)^2},
\end{equation}
where $\Gamma = 1 / \pi \tau$, and the relationship between the gigahertz frequency $\nu$ and the inverse-centimeter wavenumber is again $\sigma$ is $\nu = \sigma \times 30 \, \rm{GHz} \, \times 1 \, \rm {cm}$. 
Equation \ref{eq:analytic_bp} is then normalized to peak at a maximum of 1.
Figure \ref{fig:fts} shows the results of this process for one SPT-SLIM pixel.
The colored peaks are the detector bandpasses from Equation \ref{eq:analytic_bp}, and the gray infill shows the median atmospheric transmission profile for the South Pole summer over the SPT-SLIM frequency band.
Note that several of the lowest-frequency detectors have bandpasses that intersect with the 118 GHz oxygen line, thus experiencing a much higher atmospheric loading than other detectors.
This intentional feature is consistent among all five operational SPT-SLIM pixels.

\section{Atmospheric Loading Model}
\label{sec:atmos}

To estimate the atmospheric loading on the SPT-SLIM detectors, we make use of atmospheric opacity measurements from the 850 GHz  tipping radiometer (or ``tipper") \cite{Radford_2016} installed on the roof of the same building that houses the SPT.
We use the tipper data in combination with the \textit{am} atmospheric modeling package \cite{am_software} to generate simulated atmospheric spectra for each atmospheric response measurement taken by SPT-SLIM.
We then integrate the detector bandpasses described in the previous section over these simulated spectra to obtain an estimated atmospheric load.

\subsection{Atmospheric simulations}

\begin{figure}[t]
\includegraphics[width=6.6cm, trim={0cm, 0cm, 1cm, 1cm} ,clip]{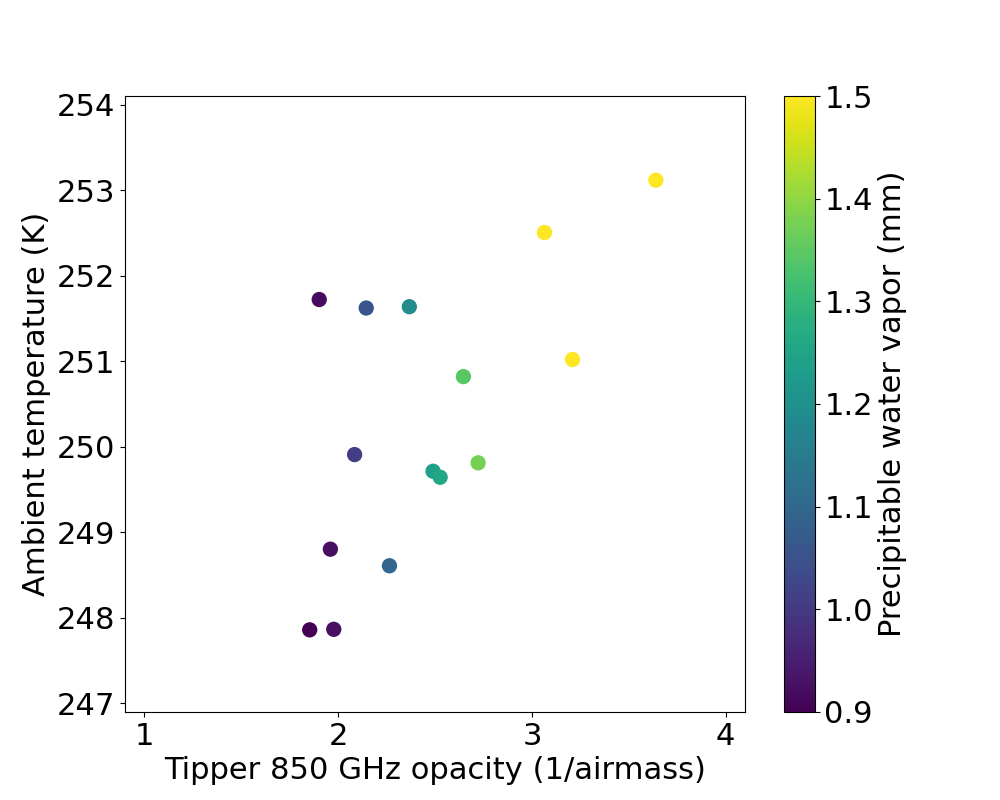}
\includegraphics[width=6.15cm, trim={0.755cm, 0.5cm, 1cm, 0.5cm} ,clip]{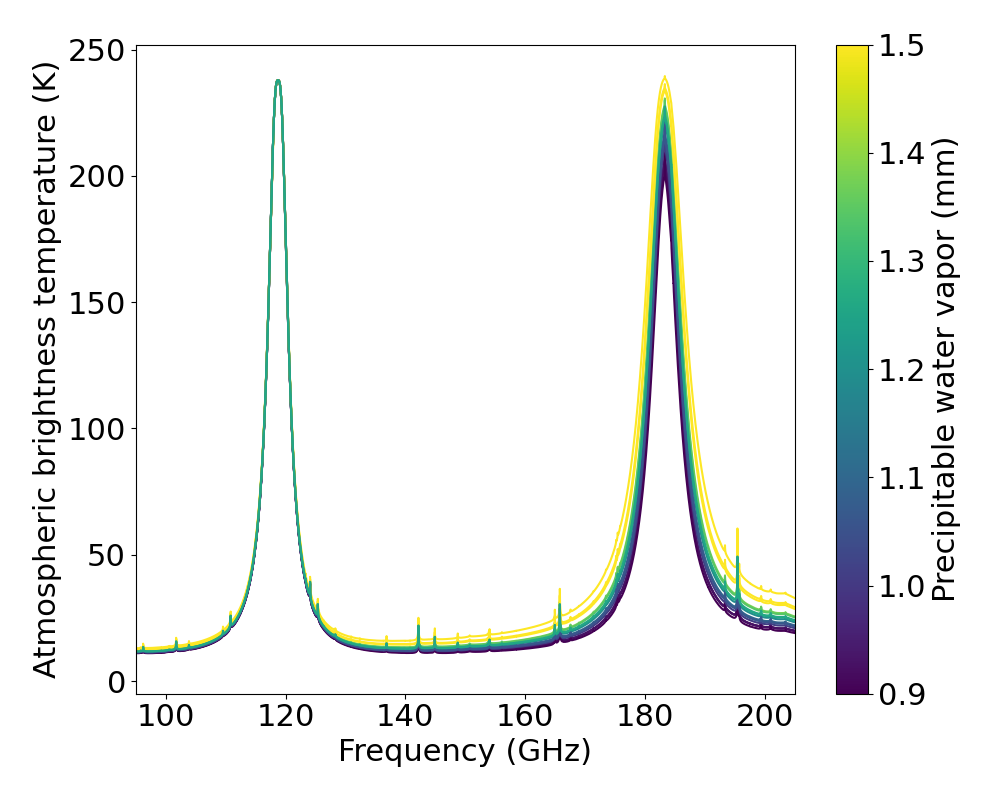}
\centering
\caption{ \textit{Upper:} 850 GHz atmospheric tipper opacity, South Pole ambient temperature, and the resulting PWV (from Equation \ref{eq:radford}) for the fifteen atmospheric response observations. \textit{Lower:} Atmospheric spectra generated by \textit{AM} for each observation. Note that only the PWV was changed for each observation, using scale factors obtained from Equation \ref{eq:scale_pwv}.}
\label{fig:atmos}
\end{figure}

The tipper took 1-3 measurements over the course of each hour-long SPT-SLIM atmospheric responsivity observation.
Each measurement generated an atmospheric opacity $\tau_{850}$, which we averaged over each SPT-SLIM observation.
When more than one tipper measurement occurred during an SPT-SLIM observation, temperatures and opacities measured by the tipper varied by less than 3\% over the course of the observation.
We then applied the following model to convert this average opacity to a PWV \cite{Radford_2016, Radford_email}:
\begin{equation}
\label{eq:radford}
{\rm PWV [mm]} = -0.2  + \tau_{850} (-0.671+ T_{\rm amb} * 0.005 ).
\end{equation}
Here $T_{\rm amb}$ is the ambient temperature obtained from the weather station located on the same building roof directly adjacent to the tipper.
To input this information into the \textit{am} software, each PWV must then be converted to a unitless ``$\rm H_2O$ scale factor" $n_{\rm scale}$, which describes the scaling of the water vapor component of the South Pole summer atmospheric profile with respect to the season median.
This is done by \cite{am_software}:
\begin{multline}
\label{eq:scale_pwv}
{\rm PWV [mm]} = n_{\rm scale} \times n_{\rm H_2O} {\rm [molecules/cm^2]} \\
 \times \frac{1}{N_A} {\rm [mol. H_2O/ molecules]}  \times 18.015 {\rm [g / mol. H_2O]}  \\
 \times 10^4 {\rm [cm^2/m^2]} \times 10^{-3} {\rm [kg/g]}.
\end{multline}
Here $n_{\rm H_2O} = 3.3 \times 10^{21} {\rm [molecules/cm^2]}$ is the column density of water vapor at the South Pole when the PWV is 1 mm, and $N_A = 6.023 \times 10^{23}$ is Avagodro's number.
Combining Equations \ref{eq:radford} and \ref{eq:scale_pwv}, each averaged 850 GHz opacity from the atmospheric tipper is converted to an $\rm H_2O$ scale factor.
We then used these scale factors in the median South Pole summer atmospheric model provided with the \textit{am} distribution to compute zenith atmospheric spectra over the SPT-SLIM observing band with a point spacing of 10 MHz.
Figure \ref{fig:atmos} shows the tipper opacity, ambient temperature, and PWV for each SPT-SLIM atmospheric response observation as well as the resulting \textit{am} atmospheric spectra.

\subsection{Optical model}
\label{sec:sidelobe}

\begin{figure}[t]
\includegraphics[width=8cm, trim={0cm, 0cm, 0cm, 0cm} ,clip]{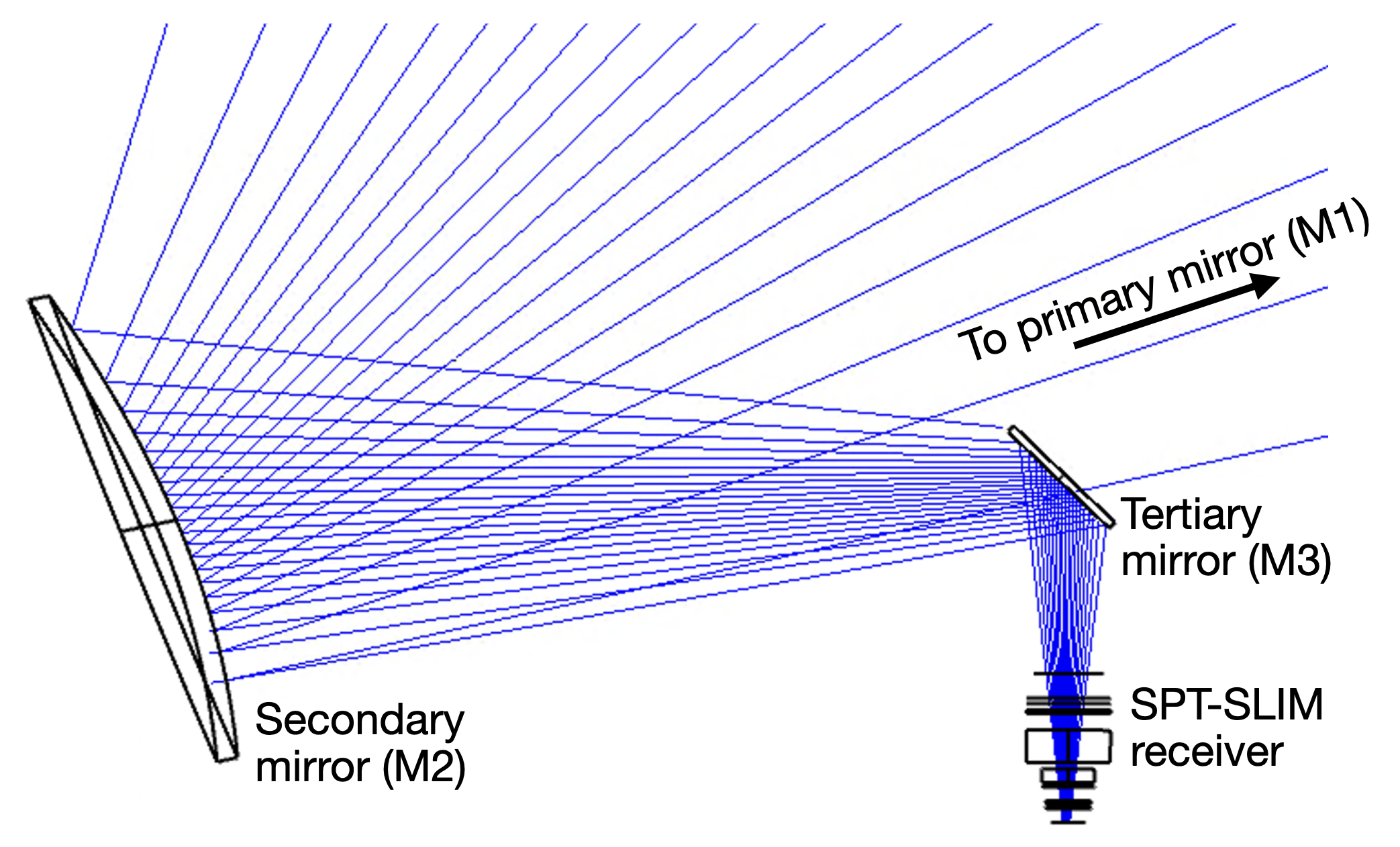}
\centering
\caption{ An optical diagram of SPT-SLIM. Light from the SPT primary mirror (M1) intersects the SPT-SLIM secondary mirror (M2), and is redirected via the tertiary mirror (M3) into the SPT-SLIM cryostat. However, a portion of the main beam intersects M3. This creates a sidelobe oriented 90 degrees offset in elevation from the main lobe.}
\label{fig:optics}
\end{figure}

SPT-SLIM is mounted in the SPT cabin, with an additional set of optics installed to redirect a portion of the main SPT beam into the SPT-SLIM receiver.
The optical path by which this is accomplished is shown in Figure \ref{fig:optics}.
Light from the SPT primary mirror (M1) intersects the SPT-SLIM secondary mirror (M2), and is redirected via the tertiary mirror (M3) into the SPT-SLIM cryostat.
However, a portion of the main beam intersects the tertiary mirror and is redirected into a sidelobe offset by $90\degree$ in elevation from the main beam.
(Equivalently, from a time-reversed perspective, the focal plane over-illuminates M3, such that not all rays reach M1.)
This sidelobe terminates on the atmosphere during the SPT-SLIM elevation response observations, creating an additional contribution to the detector loading which must be accounted for when modeling the detector response.
As the gain of the sidelobe relative to the main lobe is unknown, we introduce a fit parameter $\alpha$ to parameterize the sidelobe contribution as described below.

\subsection{Atmospheric loading computation}
\label{sec:load}

The \textit{am} simulations produce an atmospheric brightness temperature spectrum $T_b (\nu)$ and an atmospheric opacity spectrum $\tau (\nu)$.
Due to the presence of the optical sidelobe discussed in Section \ref{sec:sidelobe} we model the atmospheric temperature seen by the detectors as the sum of a main lobe in the direction that the telescope is pointed and a sidelobe that is offset in elevation by $90\degree$.
The parameter $\alpha$ below represents the gain of the sidelobe relative to that of the main lobe.
Since this value is not known, $\alpha$ is a fit parameter in our loading model:

\begin{multline}
\label{eq:tatm}
T_{\rm sky}(\nu, \theta, \alpha) =  \frac{T_0}{1 + \alpha} \Big(  \left[ 1 - \exp{(-\tau(\nu) / \sin{(\theta)})} \right] \\
+ \alpha  \left[ 1 - \exp{(-\tau(\nu) / \cos{(\theta)})} \right]  \Big).
\end{multline}
Here $\theta$ is the elevation at which the telescope is pointed, and $T_0$ is the effective zenith brightness temperature for a fully opaque atmosphere, derived from the \textit{am} output.
The total atmospheric loading on a detector is then computed using the Rayleigh-Jeans approximation $B_{\rm RJ}(T_{\rm sky})$ multiplied by each detector's bandpass $g_{\rm KID}(\nu)$ from Equation \ref{eq:analytic_bp} above:
\begin{multline}
\label{eq:psky}
P_{\rm atm}(\theta, \alpha) \\
=   \eta_{\rm opt} \int_{\nu_{\rm wg}}^{\nu_{\rm filt}} A \Omega \left[  B_{\rm RJ} ( T_{\rm sky}(\nu, \theta, \alpha)) \right] g_{\rm KID}(\nu) d\nu \\
  = \eta_{\rm opt} \int_{\nu_{\rm wg}}^{\nu_{\rm filt}} \frac{1}{2}  \frac{c^2}{\nu^2}  \left[ \frac{2 \nu^2}{c^2}  k_B T_{\rm sky}(\nu, \theta, \alpha) \right] g_{\rm KID}(\nu) d\nu \\
    = \eta_{\rm opt} k_B \int_{\nu_{\rm wg}}^{\nu_{\rm filt}}  T_{\rm sky}(\nu, \theta, \alpha)  g_{\rm KID}(\nu) d\nu .
\end{multline}
The system \'etendue is approximated by $A \Omega \approx \frac{1}{2} \lambda^2  = \frac{1}{2} \frac{c^2}{\nu^2}$ in the single-polarization beam-filling limit.
The integral is computed over the SPT-SLIM band from the waveguide cutoff frequency $\nu_{\rm wg} = 110 \, \rm GHz$ to the metal-mesh filter low-pass cutoff frequency $\nu_{\rm filt} = 210 \, \rm GHz$, and $k_B$ here is the Boltzmann constant.
The optical efficiency $\eta_{\rm opt}$ of the entire SPT-SLIM system, from the primary mirror through to the detectors, is assumed to be 50\% here, which is the maximum possible for this spectrometer design.
The value of $\eta_{\rm opt}$ for SPT-SLIM is currently unknown, and is likely to be much lower than this.
However, the difference in value of $\eta_{\rm opt}$ is absorbed into the definition of the fit parameter $A$ in Equation \ref{eq:freq_model} below, so that computing an effective brightness temperature does not require the value of $\eta_{\rm opt}$ to be known \textit{a-priori}.

\section{Detector response}
\label{sec:model}

The response of an MKID to an optical loading power $P$ is known to follow \cite{Gao_2008}:
\begin{equation}
\label{eq:resp_model}
\frac{dx}{dP} = R \frac{1}{\sqrt{P}},
\end{equation}
where $x = \frac{f-f0}{f0}$ is the fractional shift between the MKID unloaded resonant frequency $f_0$ and the loaded detector frequency $f(P)$.
Integrating this gives:
\begin{equation}
\label{eq:freq_model}
 f(P) = A \sqrt{P} + f_0,
\end{equation}
which is the model to which we fit the SPT-SLIM atmospheric response data. 
The responsivity constant in Equation \ref{eq:resp_model}, $R = \frac{A}{2f_0}$, is dependent on material and design parameters of the MKID.
The fit parameters in this model are $A$ and $f_0$, which are fit jointly with the parameter $\alpha$ in Equation \ref{eq:tatm} as described below.

\subsection{Fitting the data}

\begin{figure}[t]
\includegraphics[width=7cm, trim={0.7cm, 0.7cm, 0.7cm, 0.7cm} ,clip]{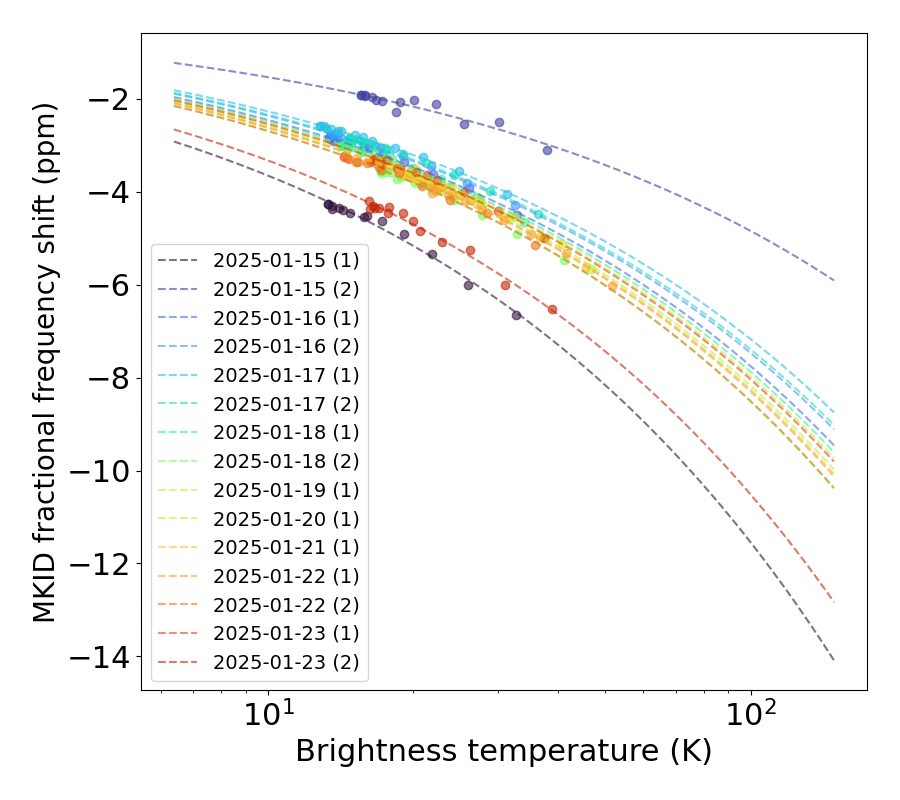}
\centering
\caption{Model fits for a detector at $\nu_0 = 154 \, \rm{GHz}$. The scattered points are measured values of MKID frequency as a function of telescope elevation, with each color corresponding to a different atmospheric response observation. Dotted lines are fits to the data of the model described in Section \ref{sec:model}.} 
\label{fig:fits}
\end{figure}

Each atmospheric response observation produces a set of MKID resonant frequencies $f(\theta)$ as a function of elevation.
These are fit with the model given by Equation \ref{eq:freq_model}, where the power $P$ is obtained from Equation \ref{eq:psky}.
The fit parameters are $\alpha$ from Equation \ref{eq:psky}, $A$ from Equation \ref{eq:freq_model}, and $f_0$ from Equation \ref{eq:freq_model}.
From these fits, we can plot the frequency response as a function of sky loading.
We convert the x-axis from power to an effective brightness temperature $T_b$, which is independent of optical efficiency:
\begin{multline}
P_{\rm RJ} (T_b) = \eta_{\rm opt} \int_{\nu_{\rm wg}}^{\nu_{\rm filt}} \frac{1}{2} \frac{c^2}{\nu^2}
 \left[ \frac{2 \nu^2}{c^2}  k_B T_b \right] g_{\rm KID} (\nu) d \nu \\
 = \eta_{\rm opt} k_B T_b \int_{\nu_{\rm wg}}^{\nu_{\rm filt}}  g_{\rm KID} (\nu) d \nu
 \equiv \eta_{\rm opt} k_B T_b G_{\rm KID}. \\
 \end{multline}
Here $G_{\rm KID}$ is the integral of the normalized detector bandpass $g_{\rm KID}(\nu)$ from Equation \ref{eq:analytic_bp}. Therefore the brightness temperature is:
\begin{equation}
\label{eq:tbrj}
T_b = \frac{P_{\rm RJ}}{\eta_{\rm opt} k_B G_{\rm KID}}.
\end{equation}
Figure \ref{fig:fits} shows fits to the data for a detector at 154 GHz.
The scatter between the different observations is likely indicative of a discrepancy between the tipper-to-\textit{am}-to-loading model and reality.
The value of $\alpha$ was found to be $\alpha = 0.22 \pm 0.16$.

\subsection{Responsivity}

\begin{figure}[t]
\includegraphics[width=7.7cm, trim={0.8cm, 0.8cm, 0.8cm, 0.8cm} ,clip]{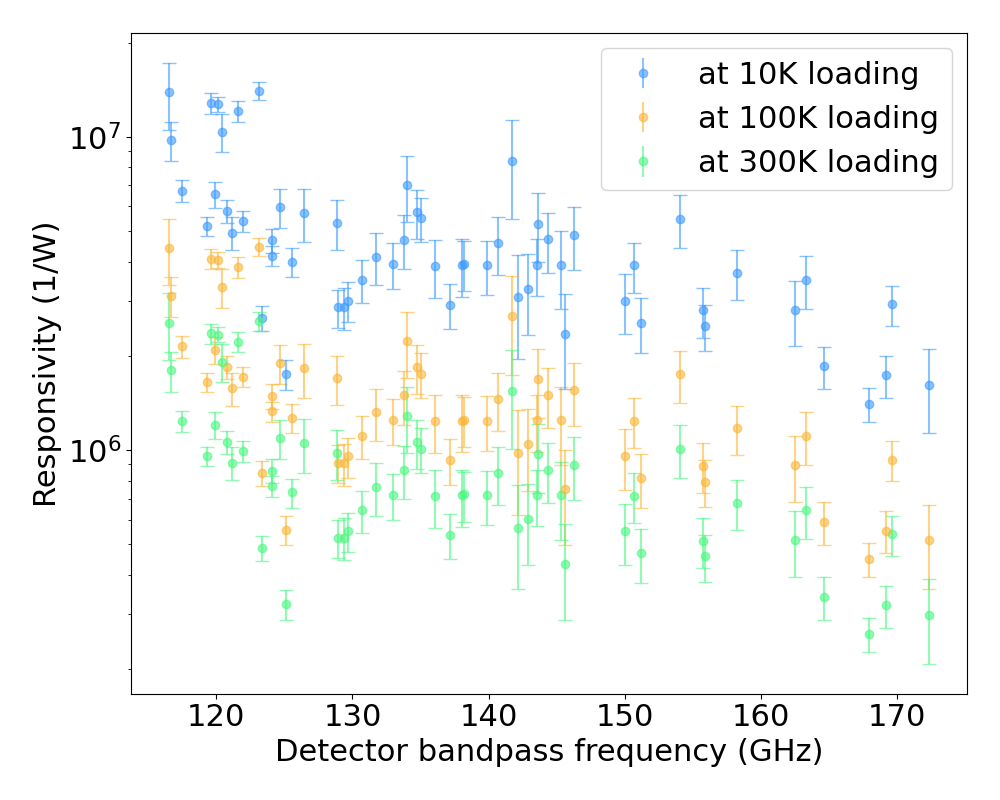}
\centering
\caption{Responsivities for one wafer of SPT-SLIM detectors as a function of detector peak bandpass frequency when loaded at brightness temperatures of 10K, 100K, and 300K. } 
\label{fig:responsivity}
\end{figure}

Evaluating our model fit parameters in Equation \ref{eq:freq_model}, we compute the detector responsivity (to power, assuming the maximum 50\% optical efficiency) as:
\begin{equation}
\label{eq:responsivity}
R_x(P) = \frac{R}{\sqrt{P}} = \frac{A}{2 f_0 \sqrt{P}}.
\end{equation}
Note that this value, unlike the brightness temperature in Equation \ref{eq:tbrj}, is not independent of per-detector optical efficiency.
Figure \ref{fig:responsivity} shows responsivity values for one SPT-SLIM pixel at brightness temperatures of 10K, 100K, and 300K as a function of detector bandpass peak frequency.
It is notable that responsivities vary across the SPT-SLIM band, with detectors at lower frequencies showing higher responsivities than those at higher frequencies.
This effect is likely due to several factors.
First, as mentioned in Section \ref{sec:load}, this analysis does not use a per-detector measurement of the optical efficiency. Variation in individual detector optical efficiencies will manifest as vertical scatter in Figure \ref{fig:responsivity}.
Second, higher frequencies of light incur greater dielectric loss as they travel down the microstrip that connects the absorptive element of the SPT-SLIM pixel to the filterbank.
The additional dielectric loss may cause the higher-frequency detectors to appear less responsive in Figure \ref{fig:responsivity}, as their loading is overestimated.
Third, the width of the detector bandpass increases as a function of frequency, meaning that higher-frequency detectors experience more loading than lower-frequency detectors at the same brightness temperature, and will therefore be less responsive.
Finally, the detectors around the 118 GHz oxygen line experience significantly more loading than other detectors, making extrapolation back to the unloaded frequency  $f_0$ less accurate.

\section{Application to Moon Observations}

\begin{table}
\caption{\label{tab:moonobs} The timestamp, moon elevation $\theta$ and moon phase $\psi$ for each SPT-SLIM moon observation.}
\begin{center}
\begin{tabular}{ c c c c }
 Date & Time (UTC) &  $\theta$ (deg) & $\psi$ (deg)\\
 \hline
20250121 & 08:18:33 & 11:20:45 & 264 \\  
20250122 & 03:51:15 & 15:25:15 & 273 \\  
20250122 & 08:35:20 & 16:21:48 & 275 \\  
20250123 & 01:45:11 & 19:31:56 & 283 \\  
\end{tabular}
\end{center}
\end{table}

SPT-SLIM observed the moon four times near the end of its observing campaign.
Table \ref{tab:moonobs} gives the timestamp, moon elevation $\theta$, and moon phase $\psi$ for each observation.

\subsection{Map processing}

Each moon scan was performed over an area of $1.8 \degree$ azimuth $\times$ $2\degree$ elevation.
The SPT-SLIM detectors were tuned immediately prior to each moon scan while pointing at an elevation close to that of the moon.
Raw data timestreams for SPT-SLIM were recorded using the RF-ICE system modified for the readout of MKIDs \cite{Rouble_2022}.
Raw timestreams are in voltage units, and are converted to MKID fractional frequency shift relative to the tuning frequency $f_t$ using the linear approximation method described in \cite{Gao_2008}.
The values necessary for this conversion are computed during the tuning at the start of each scan.
To construct maps of the time-ordered data (TOD), we follow a filtering and binning procedure typical of CMB experiments.
After conversion into units of fractional frequency shift, a 4th-order polynomial is subtracted from the TOD for each detector to remove low-frequency signals due to changes in atmospheric loading during a single scan across the source. During the fitting stage of this polynomial subtraction, the moon is masked.
The filtered TOD is then binned into a map with pixels of size 0.5 arcmin. A ``weights map" is constructed by calculating the number of TOD samples in each map bin, weighted by the average variance in the TOD segment corresponding to each raster across the source.
The map, in units of fractional frequency shift, is then estimated by dividing the binned map by the weights map.

\subsection{Calibration}
\begin{figure}
\includegraphics[width=6.5cm, trim={1.6cm, 0.7cm, 0.7cm, 0.6cm} ,clip]{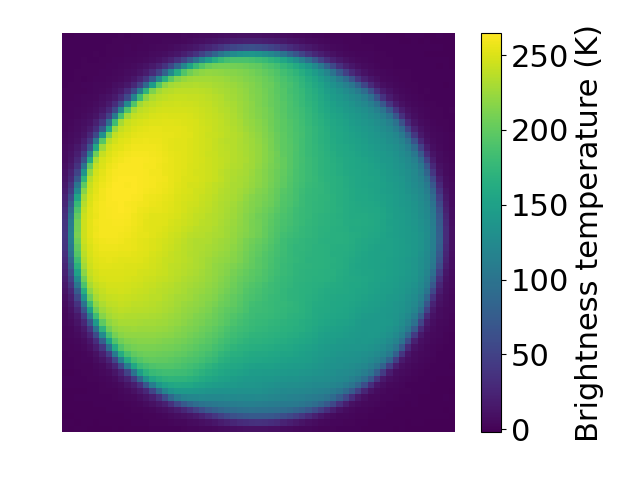}
\centering
\caption{ Example calibrated map of the moon for a detector whose bandpass peaks at 154 GHz.  This map was converted from fractional frequency shift to power by applying the calibration given by Equation \ref{eq:source_power} and was then converted to brightness temperature using Equation \ref{eq:tbrj}.} 
\label{fig:moon_maps}
\end{figure}

Using the fits to the model given in Equation \ref{eq:freq_model}, the frequency shift of a detector can be converted into an incident ``power," again assuming maximum (50\%) optical efficiency.
The telescope observes a source by pointing at the sky near the source, tuning the detectors, and then scanning over the source.
Thus, when tuning, the detectors are seeing atmospheric power $P_{\rm atm} (\theta_t)$ at the tuning elevation $\theta_t$. 
Applying the fit model from Section \ref{sec:model}, the frequency $f_t$ at which the detector is tuned can be written:
\begin{equation}
f_t = A \sqrt{P_{\rm atm} (\theta_t)} + f_0,
\end{equation}
so that
\begin{equation}
\label{eq:tuning}
 P_{\rm atm} (\theta_t)= \frac{(f_t - f_0)^2}{ A^2}.
\end{equation}
For this analysis, we assume that the atmospheric loading at the tuning elevation $P_{\rm atm}(\theta_t)$ is equal to the atmospheric loading at the source elevation $P_{\rm atm}(\theta_s)$.
This is equivalent to assuming that the atmosphere doesn't change significantly from the top to bottom of a scan. 
(For SPT-SLIM moon observations, the sky loading change over the $0.5\degree$ elevation difference between the top and bottom of the moon is roughly $2 \%$.)
When the telescope is pointed at the source, the detector is loaded by both the source and the atmosphere. Its frequency can be written:

\begin{equation}
\label{eq:source_freq}
f_s = A \sqrt{P_s + P_{\rm atm}(\theta_s)} + f_0 \approx A \sqrt{P_s + \frac{(f_t - f_0)^2}{ A^2} } + f_0,
\end{equation}
where $P_s$ is the power from the source and the second step comes from applying Equation \ref{eq:tuning}. \,
The map produced from the source observation is in units of fractional frequency shift $x = (f_s - f_t)/f_t$, referenced to the tuning frequency. Solving for $f_s$, and equating to Equation \ref{eq:source_freq}, we have:
\begin{equation}
\label{eq:source_power}
f_s = f_t(x+1) \approx A \sqrt{P_s + \frac{(f_t - f_0)^2}{ A^2} } + f_0. \\
\end{equation}
Solving for $P_s$ leads to:
\begin{equation}
\label{eq:source_power}
P_s =  \left( \frac{f_t x}{A} \right)^2 + \frac{2 f_t x (f_t - f_0)}{A^2}.
\end{equation}
The first term of Equation \ref{eq:source_power} is the raw power from the source, transmitted through the atmosphere, that lands on the detector.
The first term is always positive.
Loading from the atmosphere decreases detector responsivity, making the detector respond less dramatically to power from the source.
The second term of Equation \ref{eq:source_power} corrects for this effect.
Note that both $x$ and $f_t - f_0$ are negative, since detector frequency decreases with increased loading, and so the second term is also positive.
This power is then converted to a brightness temperature, again using Equation \ref{eq:tbrj}.
Figure \ref{fig:moon_maps} shows an example brightness temperature map of the moon after the calibration is applied.

\subsection{Moon temperature spectra}

\begin{figure}[t]
\includegraphics[width=9cm, trim={1cm, 0.3cm, 2.5cm, 1.7cm} ,clip]{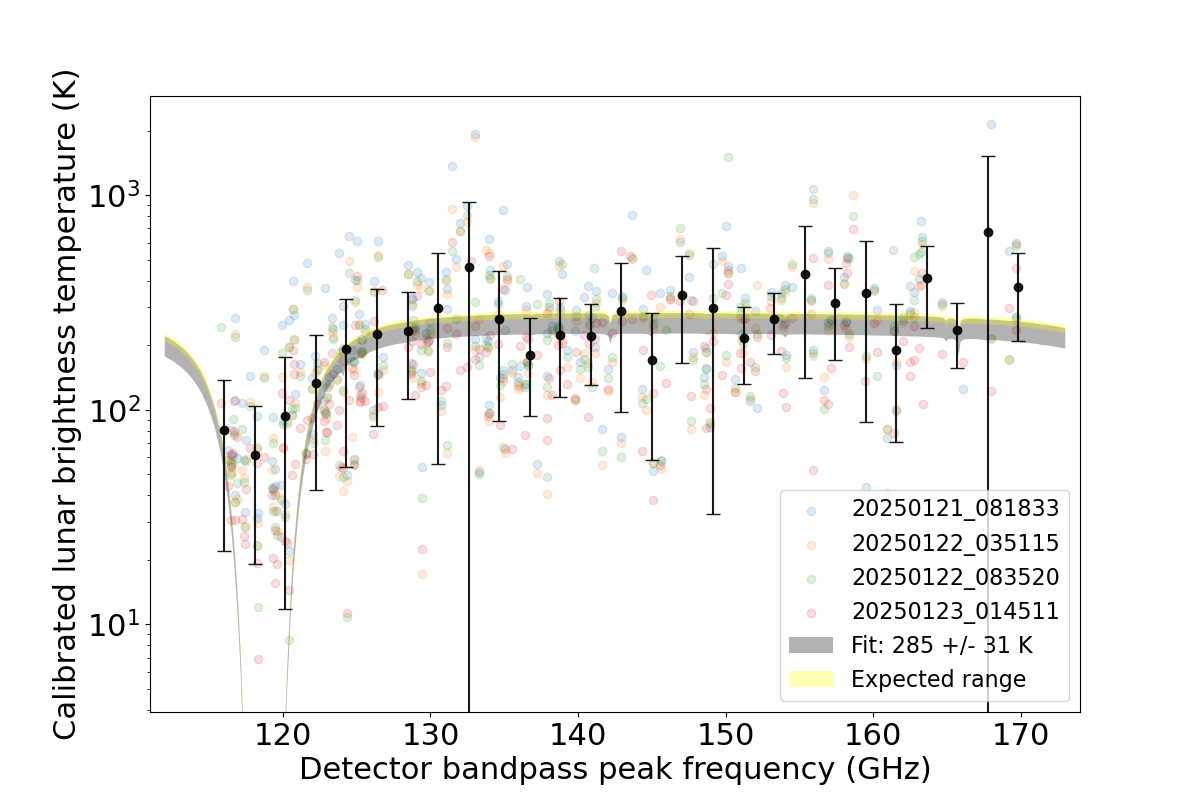}
\centering
\caption{SPT-SLIM calibrated moon spectra. Each set of colored points shows moon temperature measurements for all functional SPT-SLIM detectors for a given moon observation. Spectra are corrected for moon elevation and phase so that all four observations are comparable, as described in Equation \ref{eq:T_correction}. The combined data from all observations is binned into bins of width 2 GHz. The averages of these bins are shown in black, and the error bars are the bin standard deviations. Finally, the shaded gray region represents a fit to the binned data of a fixed-temperature moon transmitted through the mean South Pole summer atmosphere. We find a moon blackbody temperature of  $T_{\rm moon} = 285 \pm 31$K. The yellow region represents the expected moon temperature range from \cite{Sueno:2023mrj}. }
\label{fig:moon_spectra}
\end{figure}

The brightness temperature is averaged over the lunar disc to obtain an integrated moon temperature for each map.
The observed temperature of the moon depends on its phase $\psi$ as \cite{Sueno:2023mrj}:
\begin{multline}
\label{eq:moon_phase}
T_{\rm obs}(\psi, \delta) = 225 \times P(\psi, \delta) \\
= 225 \left[ 1 + \frac{0.77}{\sqrt{1 + 2 \delta + 2 \delta^2}} \cos \left( \psi -  \frac{\delta}{1+\delta}  \right) \right],
\end{multline}
where $\delta = 0.3 \lambda = 0.3 c / \nu$.
We correct for the differing moon phases across observations by multiplying the observed temperature $T_{\rm obs}$ by a phase correction so that all moon temperature data is comparable at an effective phase of $0 \degree$.
This is represented by the first ratio in Equation \ref{eq:T_correction} below.
In addition, the source power is attenuated by its transmission through the atmosphere, and the magnitude of this attenuation depends on the elevation $\theta$ of the source.
To obtain comparable temperatures between moon observations, each observation is corrected to an effective elevation of $\theta = 15 \degree$ using the ratio of the \textit{am}-simulated atmospheric transmission $\rm Tx (\theta, \nu)$ at $\theta = 15 \degree$ to the transmission at the observing elevation $\theta_{\rm obs}$.
This is the second ratio in Equation \ref{eq:T_correction}.
Thus the corrected moon temperature for each detector, for each observation, at an effective phase angle $\psi = 0\degree$ and moon elevation $\theta = 15 \degree$ is:
\begin{multline}
\label{eq:T_correction}
T_{\rm obs}(0 \degree, 15 \degree, \nu) = \\
T_{\rm obs}(\psi_{\rm obs}, \theta_{\rm obs}, \nu) \frac{P(0 \degree, \delta(\nu))}{P(\psi_{\rm obs}, \delta(\nu))} \frac{\rm{Tx}(15 \degree, \nu)}{\rm{Tx}(\theta_{\rm obs}, \nu)},
\end{multline}
where $\psi_{\rm obs}$ and $\theta_{\rm obs}$ are the lunar phase angle and elevation respectively for each observation, as shown in Table \ref{tab:moonobs}.
The corrected moon temperatures for each detector, for each moon observation are represented by the colored points in Figure \ref{fig:moon_spectra}.
The colored points are then binned (across all observations) into 30 bins of width 2 GHz.
The black points in Figure \ref{fig:moon_spectra} are the bin averages.
The error bars on the black points are the standard deviations of each bin.
Finally the binned data is fit with the following model:
\begin{equation}
\label{eq:moon_fit}
T_{\rm obs}(\psi=0 \degree, \theta=15 \degree, \nu) = T_{\rm moon} \times \rm{Tx}(\theta =15 \degree, \nu),
\end{equation}
where $\rm{Tx}(15 \degree, \nu)$ is the \textit{am}-simulated median South Pole summer atmospheric transmission at 15 \degree elevation, and the fit parameter $T_{\rm moon}$ is our ``final" measured full moon temperature.
The decreased atmospheric transmission around the oxygen line causes the observed moon temperature to also decrease around 118 GHz.
We find $T_{\rm moon} = 285 \pm 31$K, which is represented by the gray region in Figure \ref{fig:moon_spectra}.
This agrees with the expected moon temperature range from \cite{Sueno:2023mrj}, which is overplotted in yellow in Figure \ref{fig:moon_spectra}.

\section{Conclusion}

We have used measurements of the SPT-SLIM detector response to atmospheric loading to obtain a calibration which converts detector fractional frequency shift to brightness temperature.
Atmospheric loading was estimated using a combination of on-site detector bandpass measurements, on-site atmospheric measurement instruments and the simulation software \textit{am}.
The detector response to atmospheric loading was then fit with a model that includes the presence of an optical sidelobe.
From these model fits, detector responsivity estimates were derived.
The results of the model fits were then used to calibrate four observations of the moon performed by SPT-SLIM and to extract lunar brightness temperature spectra.
The agreement of the SPT-SLIM-measured lunar temperature spectrum with literature measurements provides on-sky validation of the spectral sensitivity of the SPT-SLIM receiver.
This model could be improved with additional atmospheric loading information as well as more complete measurements of the SPT-SLIM beam.
As SPT-SLIM observed other sources besides the moon, this calibration method may be applied to other source maps to extract spectra.

\section*{Acknowledgment}

SPT-SLIM is supported by the National Science Foundation under Award No. AST-2108763. The South Pole Telescope program is supported by the National Science Foundation (NSF) through awards OPP-1852617 and OPP-2332483. Work at Argonne, including use of the Center for Nanoscale Materials, an Office of Science user facility, was supported by the US Department of Energy, Office of Science, Office of Basic Energy Sciences and Office of High Energy Physics, under Contract No. DE-AC02-06CH11357. This document was prepared by the SPT-SLIM collaboration using the resources of the Fermi National Accelerator Laboratory (Fermilab), a U.S. Department of Energy, Office of Science, Office of High Energy Physics HEP User Facility. Fermilab is managed by FermiForward Discovery Group, LLC, acting under Contract No. 89243024CSC000002. The McGill team acknowledges funding from the Natural Sciences and Engineering Research Council of Canada and the Canadian Institute for Advanced Research, and the Canada Research Chairs program. This work is supported by UKRI Future Leaders Fellowship MR/W006499/1. Partial support is also provided by the Kavli Institute of Cosmological Physics at the University of Chicago. Support for this work for JZ was provided by NASA through the NASA Hubble Fellowship grant HF2-51500 awarded by the Space Telescope Science Institute, which is operated by the Association of Universities for Research in Astronomy, Inc., for NASA, under contract NAS5-26555. KSK was partially supported by an NSF Astronomy and Astrophysics Postdoctoral Fellowship under award AST-2001802, and by SLAC under award LDRD-24-008. 

We thank Heitor Mourato, Glenn Thayer, and Jose Velho in the BU Scientific Instrument Facility for assistance with mirror fabrication, and John Kovac and Miranda Eiben for assistance with anti-reflection coating optics.
We thank Simon Radford for the model used in Equation \ref{eq:radford} to convert tipper opacities to precipitable water vapors.
KRD thanks the 2025 Winter Crew of the Amundsen-Scott South Pole Station for supporting this work.

\bibliography{ltd_2025}{}
\bibliographystyle{unsrt}
%

\end{document}